\documentclass[longauth,onecolumn]{aa}  

\usepackage{graphicx}
\usepackage{txfonts}
\usepackage{hyperref}
\usepackage{gensymb}
\usepackage{natbib}

\begin{document} 

\title{Sub-arcsecond imaging with the International LOFAR Telescope: II. Completion of the LOFAR Long-Baseline Calibrator Survey}

\titlerunning{The complete LBCS survey}
\author{Neal Jackson\inst{1}, Shruti Badole\inst{1}, John Morgan\inst{2}, Rajan Chhetri\inst{2}, Kaspars Pr\={u}sis\inst{3}, Atvars Nikolajevs\inst{3}, Leah Morabito\inst{4}, Michiel Brentjens\inst{5}, Frits Sweijen\inst{6}, Marco Iacobelli\inst{5}, Emanuela Orr\`u\inst{5}, J. Sluman\inst{5}, R. Blaauw\inst{5}, H. Mulder\inst{5}, P. van Dijk\inst{5}, Sean Mooney\inst{7}, Adam Deller\inst{8}, Javier Moldon\inst{9}, J.R.  Callingham\inst{6,5}, Jeremy Harwood\inst{10}, Martin Hardcastle\inst{10}, George Heald\inst{11}, Alexander Drabent\inst{12}, J.P. McKean\inst{5,13}, A. Asgekar\inst{5,14}, I.M. Avruch\inst{5,15}, M.J. Bentum\inst{5,16}, A. Bonafede\inst{17,18,19}, W.N. Brouw\inst{13}, M. Br\"uggen\inst{19}, H.R. Butcher\inst{20}, B. Ciardi\inst{21}, A. Coolen\inst{5}, A. Corstanje\inst{22,23}, S. Damstra\inst{5}, S. Duscha\inst{5}, J. Eisl\"offel\inst{12}, H. Falcke\inst{23}, M. Garrett\inst{1}, F. de Gasperin\inst{24,25}, J.-M. Griessmeier\inst{26,27},A.W. Gunst\inst{5}, M.P. van Haarlem\inst{5}, M. Hoeft\inst{12}, A.J. van der Horst\inst{28,29}, E. J\"utte\inst{30}, L.V.E. Koopmans\inst{13}, A. Krankowski\inst{31}, P. Maat\inst{5}, G. Mann\inst{32}, G.K. Miley\inst{6}, A. Nelles\inst{33,34}, M. Norden\inst{5}, M. Paas\inst{35}, V.N. Pandey\inst{5}, M. Pandey-Pommier\inst{36}, R.F. Pizzo\inst{5}, W. Reich\inst{37}, H. Rothkaehl\inst{38}, A. Rowlinson\inst{5,39}, M. Ruiter\inst{5}, A. Shulevski\inst{6,39}, D.J. Schwarz\inst{40}, O. Smirnov\inst{41,42}, M. Tagger\inst{26}, C. Vocks\inst{32}, R.J. van Weeren\inst{6}, R. Wijers\inst{39}, O. Wucknitz\inst{37}, P. Zarka\inst{43,27}, J.A. Zensus\inst{37}, P. Zucca\inst{5}
}

\authorrunning{LBCS Authors}

\institute{University of Manchester, Jodrell Bank Centre for Astrophysics, Department of Physics \& Astronomy
\and ICRAR, Curtin University, Bentley, WA6102, Australia
\and Ventspils International Radio Astronomy Centre,  Ventspils University College, Inzenieru 101, LV-3601, Ventspils, Latvia
\and Department of Physics, University of Durham, South Road, Durham DH1 3LE
\and ASTRON, Hoogeveensedijk 4, 7990AA Dwingeloo, Netherlands
\and Sterrewacht Leiden, Niels Bohrweg 2, 2333 CA, Leiden
\and School of Physics, University College Dublin, Belfield, Dublin, Ireland
\and Swinburne University of Technology, Hawthorne, Victoria, Australia
\and Instituto de Astrof\'isica de Andaluc\'ia (IAA, CSIC), Glorieta de las Astronom\'ia, s/n, E-18008 Granada, Spain
\and Centre for Astrophysics Research, University of Hertfordshire, Hatfield AL10 9EU, UK
\and CSIRO Astronomy and Space Science, PO Box 1130, Bentley WA 6102, Australia  
\and Th\"uringer Landessternwarte, Sternwarte 5, D-07778 Tautenburg, Germany
\and Kapteyn Astronomical Institute, Groningen University, The Netherlands
\and Currently at Shell Technology Center, Bangalore, India 562149.
\and Science and Technology B.V., Delft, the Netherlands
\and Eindhoven University of Technology, De Rondom 70, 5612 AP Eindhoven, The Netherlands
\and DIFA - Universit\'a di Bologna, via Gobetti 93/2, I-40129 Bologna, Italy
\and INAF - IRA, Via Gobetti 101, I-40129 Bologna, Italy; IRA - INAF, via P. Gobetti 101, I-40129 Bologna, Italy
\and University of Hamburg, Gojenbergsweg 112, 21029 Hamburg, Germany
\and Research School of Astronomy \& Astrophysics, Mt. Stromlo Observatory, Cotter Road, Weston Creek, ACT 2611 Australia
\and Max-Planck Institute for Astrophysics, Karl-Schwarzschild-Stra{\ss}e 1, 85748 Garching, Germany
\and Astrophysical Institute, Vrije Universiteit Brussel, Pleinlaan 2, 1050 Brussels, Belgium
\and Department of Astrophysics/IMAPP, Radboud University Nijmegen, P.O. Box 9010, 6500 GL Nijmegen, The Netherlands
\and Hamburger Sternwarte, Universit\"at Hamburg, Gojenbergsweg 112, D-21029, Hamburg, Germany
\and INAF - Istituto di Radioastronomia, via P. Gobetti 101, 40129, Bologna, Italy
\and LPC2E - Universit\'{e} d'Orl\'{e}ans / CNRS, 45071 Orl\'{e}ans cedex 2, France
\and Station de Radioastronomie de Nan\c{c}ay, Observatoire de Paris, PSL Research University, CNRS, Univ. Orl\'{e}ans, OSUC, 18330 Nan\c{c}ay, France
\and Department of Physics, The George Washington University, 725 21st Street NW, Washington, DC 20052, USA 
\and Astronomy, Physics and Statistics Institute of Sciences (APSIS), The George Washington University, Washington, DC 20052, USA
\and Ruhr-University Bochum, Faculty of Physics and Astronomy, Astronomical Institute, 44780 Bochum, Germany
\and Space Radio-Diagnostics Research Centre, University of Warmia and Mazury, ul. Romana Prawochenskiego 9, 10-719 Olsztyn, Poland
\and Leibniz-Institut f{\"u}r Astrophysik Potsdam (AIP), An der Sternwarte 16, 14482 Potsdam, Germany
\and ECAP. Friedrich-Alexander-University Erlangen-Nuremberg, Erwin-Rommel-Str. 1, 91058 Erlangen, Germany
\and DESY, Platanenallee 6, 15738 Zeuthen, Germany
\and Rijksuniversiteit Groningen, Nettelbosje 1, 9747AJ Groningen, The Netherlands
\and Université Claude Bernard Lyon1, Ens de Lyon, CNRS, Centre de Recherche Astrophysique de Lyon, 43 Boulevard du 11 Novembre 1918, 69100 Villeurbanne, France \and Max-Planck-Institut f\"{u}r Radioastronomie, Auf dem H\"{u}gel 69,
53121 Bonn, Germany
\and CBK PAN , Bartycka 18 A,00-716 Warsaw, Poland
\and Anton Pannekoek Institute for Astronomy, University of Amsterdam, Postbus 94249, 1090 GE Amsterdam, The Netherlands
\and Fakult\"at f\"ur Physik, Universit\"at Bielefeld, Postfach 100131, 33501 Bielefeld, Germany
\and Department of Physics and Electronics, Rhodes University, Makhanda (Grahamstown), South Africa
\and South African Radio Astronomy Observatory, Cape Town, South Africa
\and LESIA, Observatoire de Paris, CNRS, PSL, SU, UP, Place J. Janssen, 92190 Meudon, France
 }

   \date{Received; accepted}
 
\abstract{The Low-Frequency Array (LOFAR) Long-Baseline Calibrator Survey (LBCS) was conducted between 2014 and 2019 in order to obtain a set of suitable calibrators for the LOFAR array. In this paper we present the complete survey, building on the preliminary analysis published in 2016 which covered approximately half the survey area. The final catalogue consists of 30006 observations of 24713 sources in the northern sky, selected for a combination of high low-frequency radio flux density and flat spectral index using existing surveys (WENSS, NVSS, VLSS, and MSSS). Approximately one calibrator per square degree, suitable for calibration of $\geq200$ km baselines is identified by the detection of compact flux density, for declinations north of 30$^\circ$ and away from the Galactic plane, with a considerably lower density south of this point due to relative difficulty in selecting flat-spectrum candidate sources in this area of the sky. The catalogue contains indicators of degree of correlated flux on baselines between the Dutch core and each of the international stations, involving a maximum baseline length of nearly 2000~km, for all of the observations. Use of the VLBA calibrator list, together with statistical arguments by comparison with flux densities from lower-resolution catalogues, allow us to establish a rough flux density scale for the LBCS observations, so that LBCS statistics can be used to estimate compact flux densities on scales between 300~mas and 2$^{\prime\prime}$, for sources observed in the survey. The survey is used to estimate the phase coherence time of the ionosphere for the LOFAR international baselines, with median phase coherence times of about 2 minutes varying by a few tens of percent between the shortest and longest baselines. The  LBCS can be used to assess the structures of point sources in lower-resolution surveys, with significant reductions in the degree of coherence in these sources on scales between 2$^{\prime\prime}$ and 300~mas. The LBCS survey sources show a greater incidence of compact flux density in quasars than in radio galaxies, consistent with unified schemes of radio sources. Comparison with samples of sources from interplanetary scintillation (IPS) studies with the Murchison Widefield Array (MWA) shows consistent patterns of detection of compact structure in sources observed both interferometrically with LOFAR and using IPS.
}

 \keywords{Instrumentation:interferometers -- Techniques:interferometric -- Surveys -- Galaxies:active -- Radio continuum:galaxies -- Atmospheric physics:ionosphere}

   \maketitle
%

\section{Introduction}

The  Low-Frequency Array (LOFAR) Long Baseline Calibrator Survey (LBCS) \citep{2016A&A...595A..86J} was carried out to find radio sources in the northern sky that can be used as calibrators for observations made using the longest baselines (up to $\sim$2000 km) of the International LOFAR Telescope \citep{van-haarlem13a}. Calibrators are important to correct for the effects of frequency- and time-dependent phase corrugations on the visibility data of sources, which in the case of LOFAR are introduced mainly by the effects of the Earth's ionosphere at low radio frequencies. Specifically, the ionosphere induces a time-variable phase and a delay that increases with decreasing frequency, and is generally modelled as a thin layer \citep{cohen09} in which the temporal and spatial variations of phase can be modelled and applied to the data \citep{intema09} provided that the thin-layer approximation holds \citep{2016MNRAS.459.3525M}.  In addition, the independent system clocks in the international stations induce a constant time delay per station. Since a constant time delay corresponds to a differing phase for different frequencies within the band, this results in an artificially induced gradient in the phase of the visibility in each baseline as a function of frequency. Calibrator sources should be bright enough to allow for the delay, phase, and rate variation with time to be determined. The calibrators need to be close enough in the sky to the target such that the calibrator and the target are susceptible to the same propagation effects. Ideally, they should have known structure, but must have enough structure that is coherent on all baselines to allow a good model to be determined by self-calibration. Paper I of this series \citep{paper1} discusses in greater detail the use of the data reduction pipeline, which uses sources from the LBCS survey to calibrate LOFAR long-baseline observations.

Interferometric observations are not the only way of detecting compact structure at low frequencies. It is also possible to use observations of interplanetary scintillation (IPS), which are being carried out by the Murchison Widefield Array (MWA) in the southern hemisphere at 162\,MHz \citep{2018MNRAS.473.2965M,2018MNRAS.474.4937C,2018MNRAS.479.2318C,2019MNRAS.483.1354S,2019PASA...36....2M}, and by the Pushchino telescope in the north at 111\,MHz \citep{2019ARep...63..479T,2020ARep...64..406T}. The critical scale for sources to scintillate at these frequencies is $\sim$0.3\arcsec, well matched to the resolution of interferometric observations with International LOFAR.

\begin{figure}[h]
\begin{tabular}{cc}
\includegraphics[width=9cm]{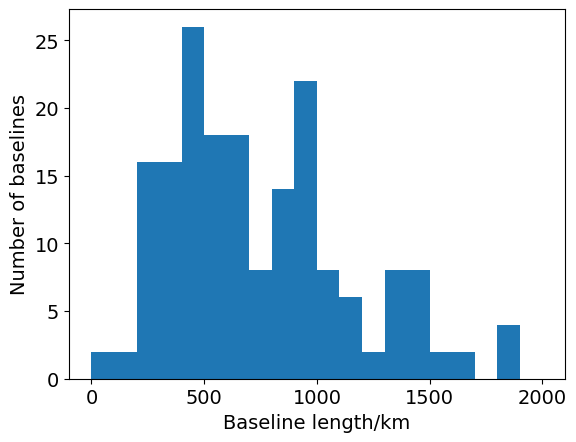}\\
\includegraphics[width=9cm]{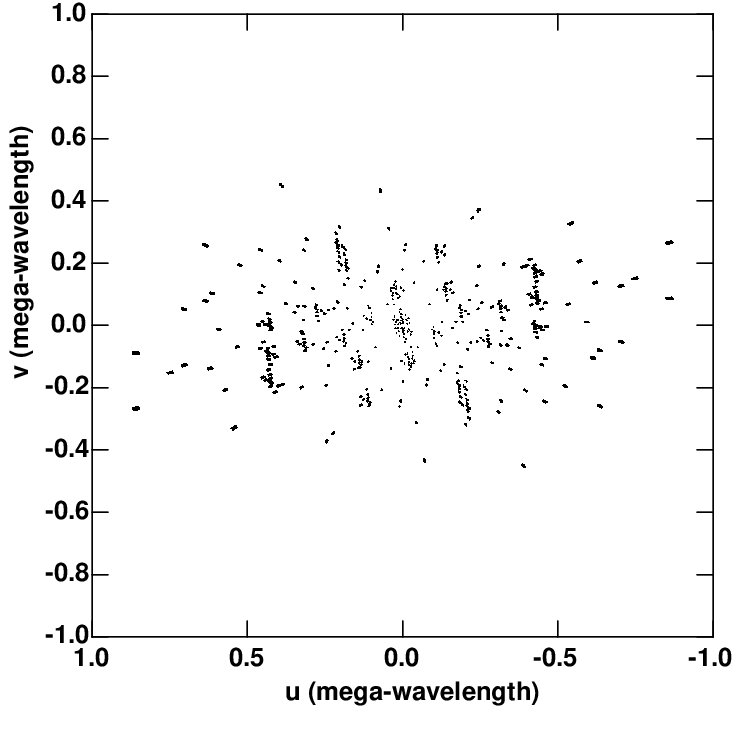}\\
\end{tabular}
\label{blength}
\caption{  Lengths and u-v coverage of the baselines used in the LOFAR observations. (top) Distribution of LOFAR baseline lengths between international stations. Core and remote LOFAR stations have baselines up to 100~km, so a gap in the $u-v$ coverage still exists between 100 and 200~km. The longest baseline, 1890~km between Birr (Ireland) and Baldy (Poland) will be exceeded soon with the advent of the station in Irbene (Latvia). (bottom) Snapshot $u-v$ plane coverage for a source at $34^{\circ}$ declination.}
\end{figure}

LOFAR consists of 40 Dutch stations, 24 of which are `core' stations located within $\sim$4~km of the centre of the array in Exloo, Netherlands. There are also 13 international LOFAR stations: 6 in Germany (Effelsberg, Unterweilenbach,  Tautenburg,  Potsdam-Bornim, Norderstedt, and J\"ulich), 3 in Poland (Borowiec, Baldy, and \L{}azy) and 1 each in France, UK, Ireland, and Sweden (Nan\c{c}ay, Chilbolton, Birr, and Onsala, respectively). A LOFAR station in Irbene, Latvia, is operational and one in Medicina, Italy, is being planned at the time of writing, although they did not take part in the LBCS observations. Figure \ref{blength} shows the distribution of the LOFAR baseline lengths between the international stations. At 140~MHz a baseline of 1300~km corresponds to a resolution of 0\farcs3, so international LOFAR is capable of routinely delivering resolving power of this order, well matched to instruments such as the VLA and e-MERLIN at gigahertz (GHz) frequencies. As a dedicated instrument with a wide field of view, it can deliver sensitivities comparable to low-frequency VLBI \citep{lenc08}, but with much greater survey speed and available observing time.

A number of calibrator surveys have been carried out for GHz frequency radio interferometers, such as the JVAS survey \citep{patnaik92} for the VLA and, for very compact sources, the VLBA calibrator list \citep{2002ApJS..141...13B} and Radio Fundamental Catalogue\footnote{astrogeo.org/rfc}. However, none of these previous surveys can reliably identify suitable calibration sources for long-baseline imaging with LOFAR.  Not all sources in densely populated arcsecond-resolution catalogues  contain components smaller than a few tenths of an arcsecond, necessary to be detected on baselines to the outlying International LOFAR stations.  On the other hand, catalogues of GHz VLBI sources are sparsely distributed (relative to the smaller isoplanatic patch size at LOFAR frequencies) and may contain sources with inverted radio spectra that are too faint at metre wavelengths \citep{2017ApJ...836..174C}.

As well as supplying assessments of suitability for array calibration, the LBCS survey also contains scientifically useful information, namely a number of measurements of correlated flux at 140 MHz frequency, on spatial scales below 1\arcsec, for a very large number  of sources ($\sim$25000). This can be used to determine, for example, the compactness of hotspots or sizes of small sources in large samples.

This paper is organised as follows. In section \ref{catalogue} we present the final catalogue, review the selection, discuss the observations and data processing, present the final catalogue, describe the coherence measures in the catalogue, and give an updated guide to its use. In this section we also use the catalogue to derive some statistics about the ionospheric phase and delay coherence at 140 MHz frequencies. Finally, in section 3 we derive some results about compact sources at 140~MHz, and we make detailed comparisons between LOFAR-interferometric and MWA IPS observations in a common field.

\section{The final catalogue}
\label{catalogue}

We briefly review the survey selection, the full details of which are given in \citet{2016A&A...595A..86J}, before updating it to present the statistics of the final catalogue and some elementary results that can be derived from it.

\subsection{Selection}
An initial pilot project for LBCS \citep{moldon15a} covered about 100 square degrees. Subsequently, the vision for the main LBCS survey was to cover the entire northern sky. The candidate sources for the region above 30$\degree$N were selected using three surveys, namely the  VLA Low-Frequency Sky Survey (VLSS) \citep{cohen07a,lane14a} at 74~MHz; the Westerbork Sky Survey (WENSS) \citep{rengelink97a} at 325 MHz; and an early, pre-publication version of the LOFAR Multifrequency Snapshot Sky Survey (MSSS) \citep{heald15} at 120-160 MHz. The TGSS ADR \citep{2017A&A...598A..78I} and, at declinations $<30^{\circ}$, the GLEAM survey, carried out at the Murchison Widefield Array \citep{2017MNRAS.464.1146H} were not available at the time of selection; the selection was also fixed before the publication of the VLSSr catalogue \citep{2014MNRAS.440..327L} which corrected VLSS fluxes by typically 10-20\% based on an improved primary beam correction. Building on the observation of \citet{moldon15a} that the detection probability of a source increases with increasing flux density, and with flattening of the low-frequency spectral index, a goodness statistic $g$ was defined with the form 
\begin{equation}
g = 2.0+\log_{10}S+2.0\alpha > 0.0955,
\end{equation}

\noindent where $S$ is the WENSS flux density in Jy, and $\alpha$ is the low-frequency spectral index, defined by $S_\nu\propto\nu^{\alpha}$. Ideally, all sources with a value of $g$ greater than some threshold should be observed. However, the r.m.s. noise level of VLSS is about 100~mJy, which is too high to be certain that all potentially compact sources detectable with LOFAR would be selected for observation. Therefore, an early draft of the MSSS catalogue at 150~MHz was also used for selection. Subsequently, the final MSSS catalogue became available. However, by that stage the observing had already begun, and the observation schedule, which contains observations of many sources at once in particular patches of sky, was fixed. 

In the part of the sky north of 30$^{\circ}$N, within the footprint of the WENSS 325 MHz catalogue, sources were selected with $g>0.0955$. A total of 13070 sources were selected using the spectral index between 74~MHz and 325~MHz using VLSS and WENSS; 919 additional sources that met the criterion were not observed  mainly because they were listed as multiple sources in WENSS, and were  less likely to have compact structure on sub-arcsecond scales. A further 5809 sources were selected because they appeared to have a relatively flat spectrum using the early MSSS draft, or because they appeared in the VLBA calibrator list, in order to maximise the use of the 30 available beams in each observation.

South of 30$^{\circ}$N WENSS is not available, and the selection procedure is described in \citet{2016A&A...595A..86J}; it essentially involves positional coincidences between VLSS sources and sources that are unresolved in the NRAO VLA Sky Survey (NVSS) at 1.4~GHz \citep{condon98a}. 5915 sources were selected in this way, leading to a total observing list of 24794 sources. A total of 30006 observations were made of 24713 separate sources;  some sources were observed more than once.

\subsection{Observations and data processing}

The observing procedure, as well as the pipeline for the data processing, is described in more detail in \citet{2016A&A...595A..86J}, so we summarise it briefly here before describing the updates to the process. The multi-beaming capabilities of LOFAR were used to observe 30 sources at once using the High-Band Array (HBA) in the DUAL-INNER mode, giving a primary beam FWHM of 3.96 degrees at 144~MHz. Each observation lasted 3 minutes with a 3 MHz bandwidth beginning at 140.16~MHz per source beam, divided into 64 channels. The core stations were phased up within the dataset after the observations, using calibrator observations, into a superstation (ST001), and into  a second superstation (ST002) using only the central six stations. Observations were conducted beginning in 2014 December and ending in 2019. The observation log is summarised in Table~\ref{observations}.

\begin{table}[]
    \centering
    \begin{tabular}{lcc} \hline
        Observation date & Sources &Stations \\ \hline
2014-12-18 &210 &1-3,5-7\\
2015-03-05 &1467&1-9\\
2015-03-18-19 &6999&1-9\\
2015-07-29-30 &5791&1-9\\
2015-11-09 &1224&1-9\\
2015-12-10 &1442&1-9\\
2016-03-08-09 &510&1-9\\
2016-04-07 &1145&1-9\\
2016-04-25-29 &2717&2-9\\
2016-05-26 &523&1-8,10-12\\
2017-03-09 &669&1-12\\
2017-04-19 &715&1-2,4-12\\
2017-05-10 &60&2-12\\
2017-10-05 &1028&1-13\\
2017-10-31 &89&1-13\\
2018-01-04 &748&1-5,7-11,13\\
2018-09-21 &197&1-11,13\\
2018-10-19 &914&1-13\\
2018-11-08 &557&2-13\\
2018-11-14-15 &112&2-13\\
2019-05-17 &382&1-13\\
2019-05-20 &604&1-13\\
2019-05-26-27 &451&1-13\\
2019-06-08 &182&1-13\\
2019-06-14 &192&1-13\\
2019-06-21 &29&1-13\\
2019-07-15 &168&1-13\\
2019-08-15-16 &53&1-5,7-13\\
2019-10-05 &34&1-2,4-12\\
2019-11-18-19 &794&2-6,8-12\\
    \end{tabular}
    \caption{Observing dates for LBCS epochs, with numbers of observed sources and the stations present for the majority of each observation. Stations are numbered from 1=DE601 (Effelsberg) to 13=IE613 (Birr).}
    \label{observations}
\end{table}

Considerable effort was made to observe at uniformly high elevation above the horizon, due to LOFAR's reduced sensitivity at lower elevations. 82\% of the observations were made at elevations of $\geq60^{\circ}$, 3\% of observations were made at elevations $\leq40^{\circ}$ (mainly the most southerly sources) and no observations were made below 34$^{\circ}$.

As in the preliminary observations, data were converted to FITS format \citep{1981A&AS...44..363W}, and all baselines to core stations (within about 4~km from the centre of the array) and the nearer remote stations were removed from the dataset\footnote{The final datasets contain the remote stations RS208, RS210, RS310, RS407, RS409, RS508, and RS509.}. The final averaged data product consisted of 64$\times$48.9 kHz channels in the frequency direction, and 90$\times$2 s time steps, for each baseline. Global fringe fitting was performed in {\sc aips} \citep{2003ASSL..285..109G}\footnote{Astronomical Image Processing System, distributed by the US National Radio Astronomy Observatory, {\tt www.nrao.edu}.} and implemented using the {\sc parseltongue}  Python interface \citep{2006ASPC..351..497K}. The pipeline used standard settings of a 500 ns delay window and 5 mHz rate window \citep{2016A&A...595A..86J} and a 6 s solution interval for the phases in the fringe fit. The scatter in the delay and phase solutions were then used to assess the strength of compact structure in the source; a strong compact source that gives a correlated signal on a particular baseline will give little scatter, whereas a weak signal gives essentially random noise in the signal. The scatter on each station's solutions was used to give designations of `P' (clear detection of correlated flux indicative of compact structure), `S' (some relatively low signal-to-noise detection of correlated flux), and `X' (little or no correlated flux) for correlated signal strength, as described in Fig. 2 of Jackson et al. (2016).

For the full dataset, the previously described procedure was modified slightly. It was found that the quality of the solution could be noticeably affected by the data weights applied to the data prior to fringe fitting. Statistically, the weight should be proportional to $\sigma^{-2}$, where $\sigma$ is the r.m.s. error. However, this rarely gives the best solution since it effectively downweights lower signal-to-noise baselines at the cost of decreasing the number of stations making effective contributions to the solution. Two other weighting schemes were used instead, a $\sigma^{-1}$ weighting and a weighting independent of $\sigma$; both weightings were applied in the fringe fit and the result that was used corresponded to that which gave the lowest scatter in the fit.

Fringe rate-and-delay maps were also produced for the full dataset, by performing a Fourier transform on a dynamic spectrum plot (as a function of frequency and time) of the complex visibility amplitude for each baseline to ST001, and then correcting the resulting geometric distortion using the correction matrix discussed in \citet{2016A&A...595A..86J}. A bright source with strong correlated signal gives a bright point at the centre of the resulting image, and other sources in the field appear in this image; the field of view of the image is larger for shorter baselines. This procedure allows us to define a signal strength parameter rather more directly than the scatter in the phase and delay solutions, and the use of the central region of the fringe rate-and-delay map guarantees that the signal comes from the source under investigation. This statistic, $S$, is defined as the sum of the six brightest  pixels in the central 10$\times$10 area of the Fourier transform of the  dynamic visibility plot, divided by the overall rms and multiplied by 4.4. The scaling is   shown in Section~\ref{sec_fluxcal} to be such that $S$ is close to a flux scale in milliJanskys for the compact structure. In general, $S$ correlates well with the P, S, or X designation, but is rather more directly related to the data. However, because of its definition in terms of the brightest pixels in the  fringe rate-and-delay map, the statistic saturates at low fluxes; non-detections and very weak detections have $100<S<150$.

\subsection{Information in the catalogue}
\label{sec:catinfo}

The information in the results catalogue allows users to perform cone searches to find signal-to-noise data. These data products are available using a cone search on {\tt http://www.lofar-surveys.org/lbcs.html} . The repository also contains other metadata, which allows judgements to be made about the suitability of observed sources as phase calibrators. This information is used when selecting the best in-field calibrator as described in Morabito et al. (2021). The first five columns give basic data: the index name (`L' followed by a six-digit number) corresponding to the number in the observing catalogue, the right ascension and declination (taken from the WENSS catalogue), and the date and time of the observation.

\begin{table*}[]
    \centering \small
\begin{verbatim}
L326144  00:00:00.99  68:10:03.0  2015-03-19  11:39:55  XXX-XXXXX----  O  000-00000----  43   0.004125    68.167500
L450116  00:00:08.95  75:40:14.1  2016-04-07  11:26:12  XSXXPXPXP----  O  077194869----  46   0.037292    75.670583
L398791  00:00:37.59  26:12:14.9  2015-07-30  02:26:52  XXXXXXXXX----  X  000000000----   0   0.156625    26.204139
L269677  00:00:41.62  39:18:03.5  2015-03-05  12:38:02  XXXXXXXXP----  O  310220025----  56   0.173417    39.300972
L269693  00:00:42.39  35:57:41.6  2015-03-05  12:38:02  PPPPPPXPP----  A  959595275----  56   0.176625    35.961556
L410226  00:00:46.92  11:14:29.0  2015-11-09  19:30:06  XXXXXXXXX----  O  900099909----  30     0.1955    11.241389
L397819  00:00:49.47  32:55:47.7  2015-07-30  02:51:00  SSPSXPSSX----  O  959829591----  41   0.206125    32.929917
L269863  00:00:51.24  51:57:20.2  2015-03-05  12:44:02  PPPPPXPPP----  O  999991499----  53     0.2135    51.955611
L269321  00:00:53.12  40:54:01.5  2015-03-05  12:26:02  PPPPPPPPP----  O  999999999----  70   0.221333    40.900417
L269649  00:00:54.52  38:02:45.0  2015-03-05  12:38:02  PXPXPXXXP----  A  915190108----  56   0.227167    38.045833
L269313  00:01:01.52  41:49:29.2  2015-03-05  12:26:02  PPPPPSPPP----  O  999996999----  70   0.256333    41.824778
L410220  00:01:02.32  10:35:49.6  2015-11-09  19:30:06  XXXXXXXXX----  O  009900009----  30   0.259667    10.597111

\end{verbatim}

    \caption{First 12 lines of the catalogue. The columns are, in order [1] the LBCS name derived from the observing system ID of the source observation; [2] the right ascension (J2000); [3] the declination (J2000); [4] the observing date; [5] the observing time; [6] a string of letters for each telescope in numerical order from DE601/Effelsberg to IE613/Birr, where P indicates strong correlated flux as measured by the delay and phase solutions to that antenna, S marginal correlated flux, X little or no correlated flux, and a hyphen that the telescope did not participate in the observations; [7]  the quality flags, where  O is a good observation, A   an observation with anomalous amplitudes (but for which the flux indicators should still be reliable), M   a failed observation with few channels, X   a failed observation with few time steps, and Z   an observation with more than half the data flagged; [8] the $n_S$ statistic for each station, based on signal-to-noise (see text); [9] a quality flag based on the group of observations of which the source is a part (see Jackson et al. 2016); and [10,11] the right ascension and declination in decimal form.}
    \label{lbcstable}
\end{table*}

The sixth column, as in the preliminary catalogue, is a string of 13 characters, each representing the correlated flux determined from the global fringe fit on each antenna. The characters in the string are given in order of telescope name, the first being DE601 (Effelsberg) and the last  being IE613 (Birr). The seventh column provides an additional quality flag based on the data itself: `O' for a good observation; `A' for an observation with out-of-range amplitudes, defined as high points above 15 times the median amplitude; `M' and `X' for failed observations with very few channels and time steps, respectively; and `Z' for a dataset with more than half of the data flagged as bad. The eighth column, in the same format as the sixth, presents the FT signal-to-noise statistic $S$ in the form of numbers from 0 to 9 for each station. Each number $n_S$ is a code related to the signal-to-noise statistic $S$ by ${\tt int}((S-132)/26.4)$, with values below $S=132$ represented as 0 and above $S=370$ represented as 9. The ninth column, as in \citet{2016A&A...595A..86J} is a quality flag based on the proportion of detected sources in the 30-source observation. This statistic should be used with caution, as it is likely to be systematically lower in areas of the sky where selection is less efficient.

\begin{figure}[h]
\includegraphics[width=9cm]{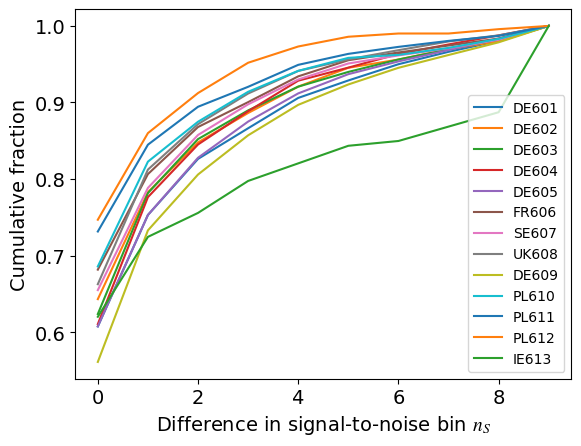}
\caption{Reproducibility of the observations. For each station, the fraction of duplicated observations which are consistent within a given difference in the FT signal-to-noise statistic is shown as a function of the difference. The reproducibility is relatively good, normally 80\% within a difference of 1 in the numerical value of the parameter.}
\label{reproducibility}
\end{figure}

A consolidated catalogue is also available as a raw-text data product, in which duplicate observations have been combined together; the first 12 lines of this catalogue are given in Table~\ref{lbcstable}. In this catalogue the highest signal-to-noise number and the best scatter within the duplicated observations are used for the assessment of each station. Duplicated observations also allow us to assess the reproducibility of the catalogue (Fig.~\ref{reproducibility}). In this figure we show the difference in the FT signal-to-noise statistic as a cumulative fraction against the difference in the statistic between different observations. For most stations over 80\% of measurements are consistent within a difference of 1 in the statistic; the exception is IE613, which was included in relatively few observations.

Other data products have been described in \citet{2016A&A...595A..86J} and have the same format. In summary, these comprise the fringe rate-and-delay map; the png file with the dynamic phase map, its Fourier transform, and the fringe solutions; and the pickle file containing a summary of all the numerical data derived from the observation. \footnote{These data products are also available using a cone search at {\tt http://www.lofar-surveys.org/lbcs.html}. The pickle file can be read using the Python/numpy routine {\tt numpy.load}, using {\tt encoding='latin1',allow\_pickle=True}. The parameter $S$ can be obtained for each antenna from the array with the key 'fftsn' by multiplying by 4.4.}

Although experience is still being accumulated, calibrators with P status for any baseline have always yielded good solutions in test fields, and calibrators without P status but with a reasonable ($n_S>3$) signal-to-noise statistic have usually given good results.

\subsection{Calibrator density}

\begin{figure}[h]
\begin{tabular}{c}
\includegraphics[width=9.2cm]{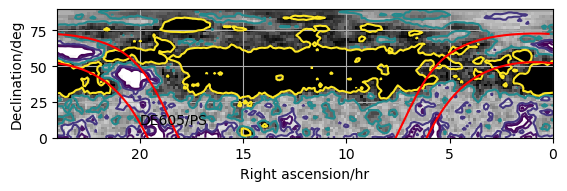}\\
\includegraphics[width=9.2cm]{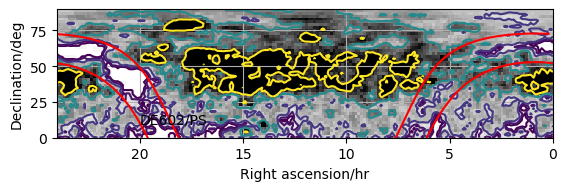}\\
\includegraphics[width=9.2cm]{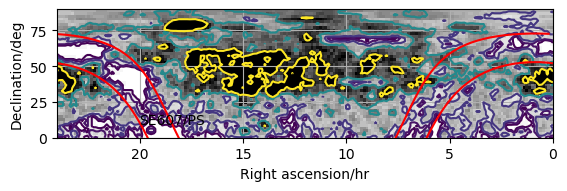}\\
\end{tabular}
\caption{Average calibrator density, considering sources graded P or S and with a smoothing kernel of 3$^{\circ}$, for three different stations: (top) DE605/J\"ulich, Germany, about 220~km from the core station at Exloo, Netherlands; (middle) DE602/Unterweilenbach, near Munich, Germany, about 580~km; (bottom) SE607/Onsala, Sweden, about 600~km. Contours are at 0.05, 0.1, 0.2, 0.5, and 1.0 sources per square degree, with the yellow contour at the highest density. The red lines indicate Galactic latitudes of $\pm 10^{\circ}$, and the two large gaps in coverage are close to the bright radio sources Cyg A and Cas A. Some small apparent gaps in coverage at higher Galactic latitudes are due to non-participation of the relevant station in one epoch of observations.}
\label{density}
\end{figure}

\begin{figure}[h]
\includegraphics[width=9cm]{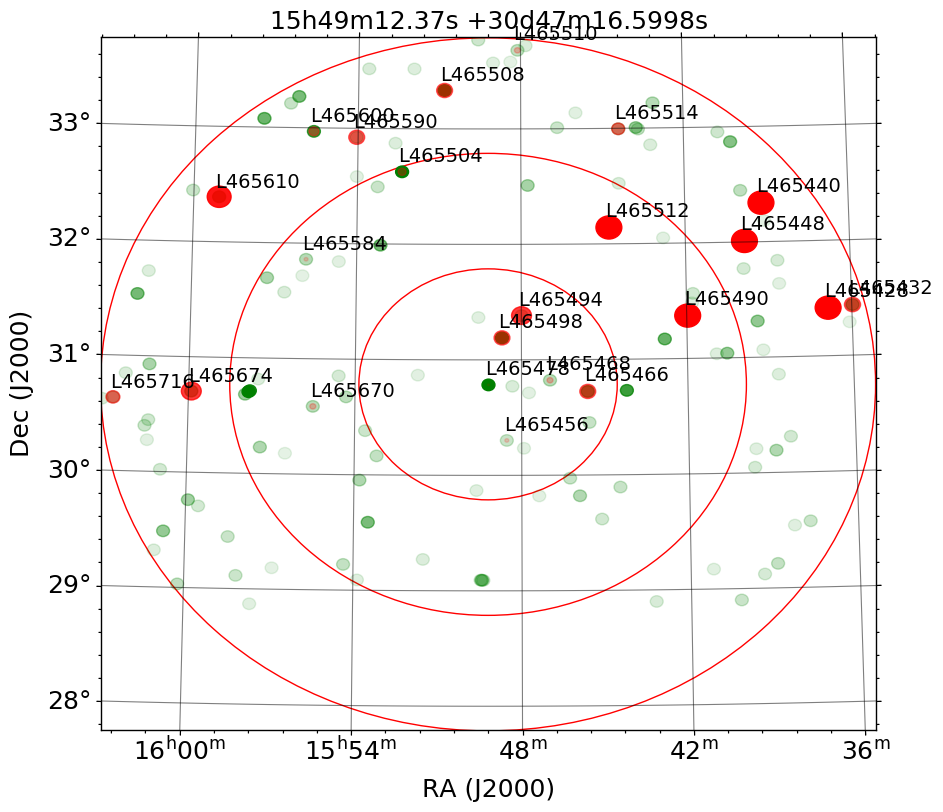}
\caption{LBCS sources in a typical field (around the gravitational lens MG~1549+3047), with the decrease in density of sources below declination 30$^{\circ}$. Red points are LBCS sources, the size indicating the fraction of stations to which significant signal is detected; green points are WENSS sources, with hue dependent on flux density.}
\label{lbcs_1549}
\end{figure}

The ideal density of suitable sources in a calibrator survey is at least one per isoplanatic patch, across which the ionospheric phase and delay effects are approximately constant. Experience with long-baseline calibration in moderate to good ionospheric conditions using the main LOFAR international-baseline test field \citep{paper1} suggests that phase solutions are reasonably transferable over about one degree. Figure~\ref{density} shows the density of sources with significant correlated flux, smoothed with a $3^{\circ}$ smoothing radius. A calibrator density of 1 per square degree is reached for all stations, including more the distant ones,  within the northern galactic cap, but decreases quickly below declinations of 30$^{\circ}$N due to the lower efficiency of selection at those declinations (Fig.~\ref{lbcs_1549}). It is also likely that the reduced sensitivity of LOFAR baselines at more southerly declinations impacts the rate of LBCS detection. For example, the fraction of well-detected P sources increases from about 35\% of the sample at declination zero to just under 50\% at  declination 28$^{\circ}$ on the DE601/Effelsberg station and from 10\% to 35\% on UK608/Chilbolton. The calibrator density is lower by a factor of 2 close to the Galactic plane, and is close to zero in a ten-degree radius from the bright radio sources Cas A and Cyg A.

In principle, delay and phase calibration can be done using a calibrator close to the target which has significant compact structure to the stations being calibrated. Sometimes, particularly at more southern declinations, the nearest compact source in the observing list is some distance from the target, and in this case it may be useful to look at the  fringe rate-and-delay map of nearby sources, which is also available as a data product, to determine whether other nearby sources have compact flux density, and thus appear on the map.

\subsection{Ionospheric coherence times as a function of baseline}

\begin{figure}[h]
\includegraphics[width=9.5cm]{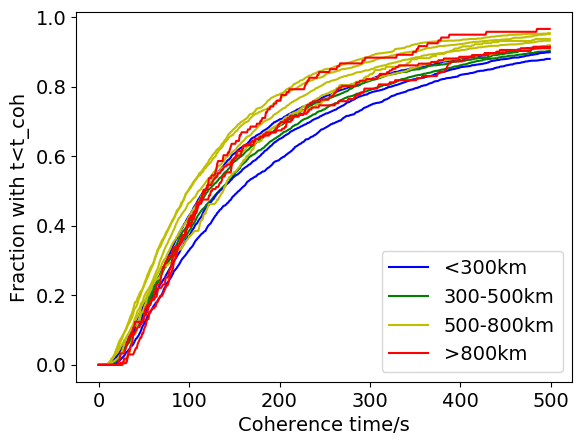}
\caption{Coherence times for each station, plotted as a cumulative fractional frequency of coherence time. Stations are colour-coded according to their distance from the core station ST001. }
\label{times}
\end{figure}

We repeat the analysis of ionospheric coherence times as a function of baseline length carried out by Jackson et al. (2016), but now with the full dataset. The coherence time is again calculated using the phase solutions from the global fringe fit, to which a second-order polynomial is fitted; the average gradient of this fit can be used to calculate a coherence time as the time needed for a phase slope of this gradient to reach one radian. The core superstation ST001 is used as the reference, and a cumulative fraction plot of coherence time for each station is shown in Fig.~\ref{times}.

For all stations the coherence time for a 50\% chance of coherence, with respect to ST001, is between 100 and 150 seconds. Stations closer to the core  generally have  longer coherence times, although the difference is not large. 

Cohen \& R\"ottgering (2009) investigated the differential atmospheric refraction between pairs of sources at different separations. They found that the position shifts arising from differential refraction, although generally increasing, only increased slowly for separations $>8^{\circ}$, corresponding to about 50~km in horizontal distance between cut points of the lines of sight in the ionosphere, at the 450 km height of the ionosphere. Figure~\ref{times} gives a measure of correlation between ionospheric phase fluctuations at cut points more widely separated. Because the separation of these points is approximately equal to the baseline lengths for a single source, it implies that some level of correlation is present between cut points separated by 250~km that is not present at 1000~km separation, and that hence the ionosphere retains some correlated phase behaviour on scales of a few hundred kilometres.

\subsection{LBCS flux density calibration}
\label{sec_fluxcal}

Although the LBCS survey does not have good {a priori} flux calibration, we can estimate the flux density scale of LBCS by reference to known sources. Because catalogued sources typically have low-frequency flux densities measured using low-resolution surveys (typically between 20$^{\prime\prime}$ and 1$^{\prime}$ resolution) we cannot use published fluxes directly for any source. Instead, we use sources that appear in the VLBA calibrator list \citep{2002ApJS..141...13B,2005AJ....129.1163P,2006AJ....131.1872P}, and are therefore likely to have all flux concentrated in a compact region. For these sources we can use flux densities in low-resolution catalogues to infer the flux density that will be seen in LBCS. In principle, noise will be added due to variability between the epoch of the catalogue and those of the LBCS observations, which is typically up to a few tens of percent at low radio frequencies (e.g. \citealt{1983A&A...118..171F}).

\begin{figure}[h]
\begin{tabular}{cc}
\includegraphics[width=9cm]{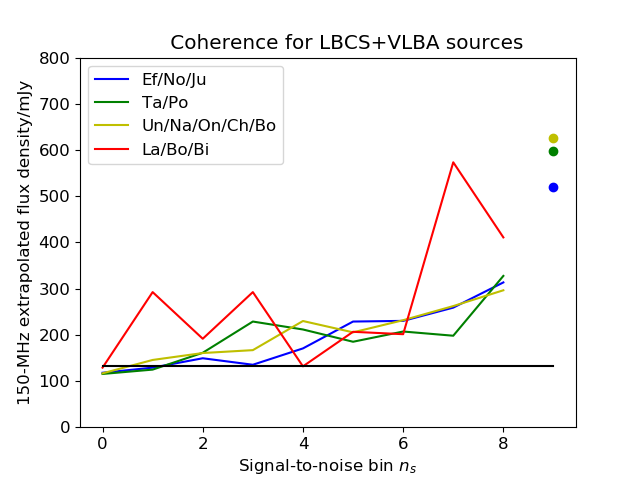}\\
\includegraphics[width=9cm]{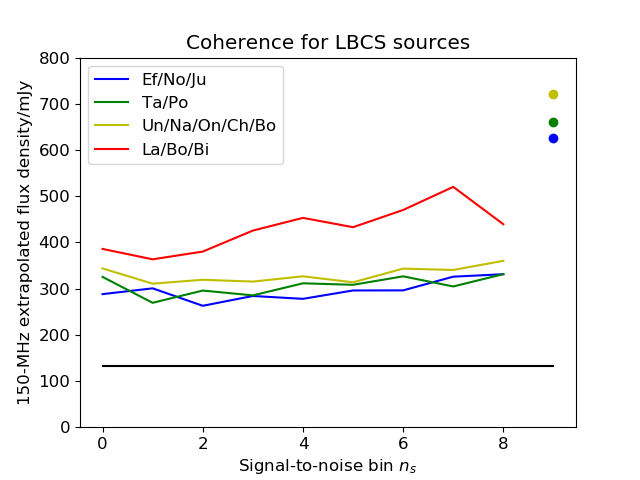}\\
\end{tabular}
\caption{  Extrapolated 150 MHz flux densities vs the LBCS signal-to-noise statistic $n_S$. Top: Signal-to-noise bin $n_S$ (with the adopted scaling of $\leq$132~mJy for $n_S=0$ to  $\geq$370mJy for $n_S=9$) vs flux for LBCS sources in the VLBA calibrator list. Blue lines (for $n_S<9$) and points ($n_S=9$) represent stations close to the Exloo core: DE601/Effelsberg, DE605/J\"ulich and DE609/Norderstedt; green for somewhat more distant stations: DE603/Tautenberg, DE604/Potsdam; yellow for more distant stations DE602/Unterweilenbach, FR606/Nan\c{c}ay, SE608/Onsala, UK608/Chilbolton, PL610/Borowiec. The red lines represent the furthest stations (PL611/\L{}azy, PL612/Baldy, IE613/Birr) and are noisy due to the small number of observations. For these stations the points for $n_S=9$ (not shown) are above the top of the plot. Bottom: As above, but for all sources, not just those in the VLBA calibrator list. The black line is the level registered by observations of blank sky, for which the signal-to-noise statistic saturates.}
\label{coh}
\end{figure}

Figure~\ref{coh} shows the results of this exercise. For each source within the footprint of both the FIRST \citep{becker95a} and WENSS surveys, we calculate its inferred 140 MHz flux density using an extrapolation from the combination of the FIRST and WENSS flux densities. We then take the median flux density calculated, in each bin of sources with a given $n_S$, and hence a given range of signal-to-noise statistic $S$, in two cases: sources that are also VLBA calibrators, and all sources. We expect the numbers in the first category to be the same for all stations since point sources should give equal signal on all baselines. This is generally true, although the curves for the furthest stations (PL611 Baldy, PL612 \L{}azy and IE613 Birr) are very noisy because they came online relatively late in the LBCS observing period, and hence relatively few sources were observed using these stations. The trend is generally linear between $1\leq n_S<9$, with $n_S=1$ corresponding to 160~mJy, and $n_S=9$ corresponding to $>$370~mJy of correlated flux (the jump at $n_S=9$ is due to all sources $>370$~mJy being included in this bin). Since $n_S=1$ corresponds to $S=160$ and $n_S=9$ corresponds to $S\geq370$, this provides the basis for the approximate identification of $S$ with the correlated flux density, in mJy, on the compactness scale corresponding to the resolution of the baseline between ST001 and any station.

If we consider all LBCS sources  (bottom panel of Fig.~\ref{coh}), the flux densities extrapolated from the FIRST and WENSS surveys are, unsurprisingly, greater for any value of the LBCS signal-to-noise statistic. This indicates that a large fraction of LBCS sources have a significant component of flux that is contributed by larger size scales than the 0\farcs3 of the international baselines, and that the median contribution of larger-scale flux density is of the order of 30-50\%.

\begin{figure*}[h]
\includegraphics[width=16cm]{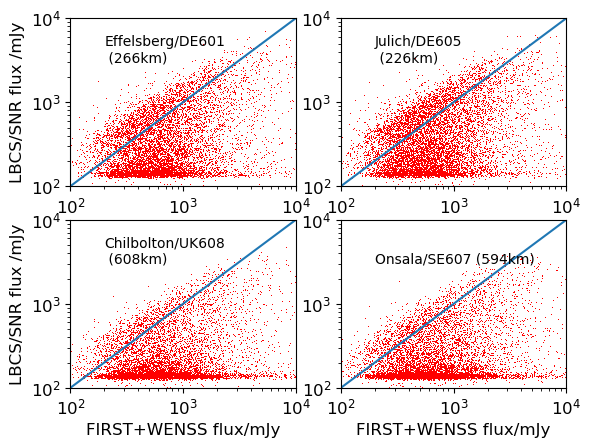}
\caption{Inferred flux from the LBCS survey measured by the signal-to-noise statistic $S$. This is plotted against inferred flux from power-law extrapolation between FIRST and WENSS flux densities for four different international stations. LBCS flux densities inferred in this way are unreliable below the lower saturation level of about 150mJy. Above this level most LBCS flux densities are (unsurprisingly) lower than the corresponding flux densities from lower-resolution surveys.}
\label{fluxcalib}
\end{figure*}

We can also plot the LBCS flux density inferred by identifying $S$ with flux density in mJy, against the extrapolated flux density from the FIRST and WENSS surveys directly. If the flux scale derived from Fig.~\ref{coh} is correct, unresolved sources should have identical flux density on both of these estimations. The results are shown in Fig.~\ref{fluxcalib}. The line of equality in this plot is an upper limit to the LBCS flux density, since large-scale structure will be resolved out by LBCS. The form of these plots, with largely unresolved sources clustering close to the line of equality and a long tail of resolved sources to the right of the line, is as expected if the flux density scale is approximately correct.

Two things are apparent from Fig.~\ref{fluxcalib}. The first is that sources with compact flux density below 200~mJy will not be reliably detected by LBCS. The second is that the upper envelope of sources in this diagram lie slightly above the line of equality for the less distant stations (DE601/Effelsberg and DE605/J\"ulich). One interpretation of this is that the identification of $S$ with flux density in mJy gives a slight overestimate of the actual flux density; the second interpretation is that the extrapolated flux density from FIRST and WENSS is a slight underestimate of the true flux density. The second is less likely as it would imply that unresolved sources steepen at lower frequencies;  we actually might expect that the extrapolation gives an overestimate of the true flux density at 140~MHz since more flux density is likely to be resolved by FIRST than by WENSS.

\subsection{Signal-to-noise reproducibility in the LoTSS survey}

We investigated the signal-to-noise ratio of sources in the field of MG~1549+3047 using both LBCS data and data obtained separately as part of project LC9-012. The LBCS fluxes for each source are from a pointing in the direction of the source, and the LC9-012 fluxes are at one central pointing; therefore, the LBCS fluxes will not be affected by bandwidth and integration time smearing, whereas the LC9-012 fluxes will be, allowing estimates of the amplitude reduction by smearing as a function of distance from a pointing centre. There are six LBCS sources in this field, one in the centre and five others at distances of up to $2^{\circ}$ from the field centre. Data processed using the LOFAR-VLBI pipeline \citep{paper1} were compared with the LBCS data by degrading the LBCS resolution by a factor of 4 in integration time and 2 in frequency to the 97 kHz/channel and 8 s integrations of the post-{\tt Split-Directions} stage of the pipeline.  The sensitivity of the pipelined data decreases by a factor of approximately 6 on a 250 km baseline when moving by about 1 degree from the field centre. This is attributed to a combination of bandwidth smearing, integration time smearing and primary beam effects, together with the field of view of ST001. We are therefore confident that the quality statistics in LBCS can be used to predict the expected quality of calibration for LBCS calibrators in standard observations, modulo ionospheric conditions.

\section{Source statistics in the LBCS sample}
\label{sources}

\subsection{Point sources from FIRST}

\begin{figure}[h]
\includegraphics[width=10cm]{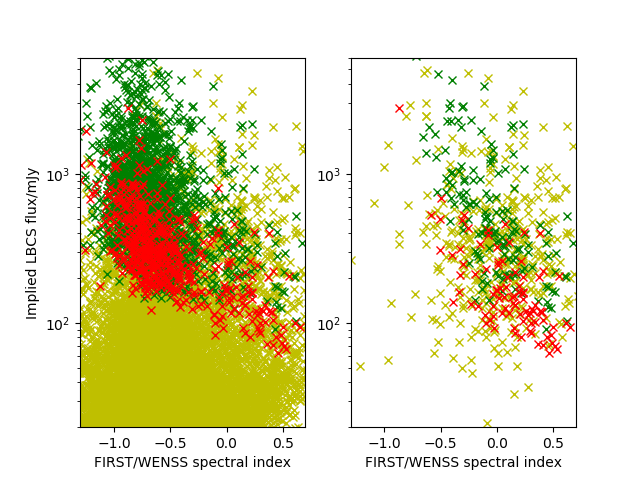}
\caption{Flux density for LBCS sources  coincident with an unresolved FIRST source and a WENSS source vs   FIRST/WENSS spectral index. The light green crosses (in background) indicate the general population of WENSS/FIRST coincidences, dark green crosses represent those sources observed with LBCS and detected, and   red crosses represent the sources observed with LBCS but not detected. The right panel contains only sources that are also VLBA calibrators.}
\label{figpoints}
\end{figure}

The FIRST survey \citep{becker95a} is a survey of a large area of the northern sky carried out at an observing wavelength of 20~cm using the VLA in B-configuration, with a maximum baseline of 12~km and a consequent resolution of about 4--5 arcseconds. In the area of overlap with LBCS and WENSS, a sample of all sources unresolved by FIRST was selected, which therefore have an intrinsic size less than 5 arcseconds. In Fig.~\ref{figpoints} are plotted the 140 MHz flux densities for these sources, extrapolated from the FIRST 1.4 GHz and WENSS 325 MHz flux densities, the latter corrected to the \cite{2012MNRAS.423L..30S} flux scale, against the FIRST-WENSS spectral index. This exercise illustrates both the bias in the LBCS selection, and the sensitivity limits of the LBCS survey. The selection by the $g$ parameter favours objects with a high flux density and  flatter spectral index. Within this selected region, we again find that the sources  with a significant correlated flux are also predominantly bright and less steep-spectrum.

The sensitivity limit of the survey can be calibrated as before by using the VLBA calibrator list (Fig.~\ref{figpoints}, right panel). The sources coincident between the LBCS and the VLBA calibrator list are a representative sample of VLBA sources both in flux density and spectral index. Again we see that sources of 140 MHz implied flux density lower than $\sim$200~mJy are undetected because they fall below the sensitivity limit of the survey, and not because they do not have a large enough percentage of their flux density on small angular scales. The majority of the sources above this limit, if they are not detected, have relatively steep spectral indices, as expected.

\label{sources}

\begin{figure*}
\includegraphics{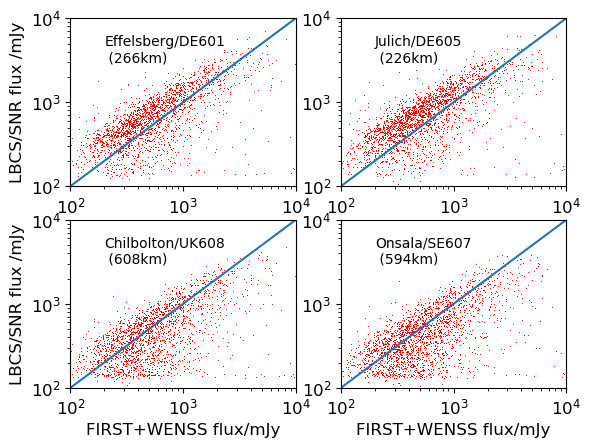}
\caption{As for Fig.~\ref{fluxcalib}, but for unresolved FIRST sources}
\label{fluxcalib_point}
\end{figure*}

Figure ~\ref{fluxcalib_point} presents the same plot as Fig.~\ref{fluxcalib}, but for the unresolved FIRST sources. This figure makes it apparent that the majority of the FIRST-unresolved sources (i.e. $<5^{\prime\prime}$)  preserve most of the flux density on 200 km baselines (to Effelsberg and J\"ulich), corresponding to about $2^{\prime\prime}$ resolution. At 300 mas resolution at least 50\% of them have lost a significant proportion of their correlated flux density. This result suggests that the predominant population, even given the LBCS selection, consists of compact steep-spectrum (CSS) sources \citep{2020arXiv200902750O} rather than classical point sources with sizes limited by synchrotron self-absorption. It also suggests that careful studies of calibrator densities will need to be undertaken before extending low-frequency arrays such as LOFAR to still longer baselines.

\subsection{Hotspots and 3CRR sources}

 The revised Third Cambridge Catalogue of Radio Sources (3CRR) \citep{1983MNRAS.204..151L} contains those extragalactic sources that have a  178 MHz flux density (KPW scale) $\geq $10 Jy, declination $\geq$ 10$\degree$, and Galactic latitude $\geq$ 10$\degree$ or $\leq$ -10$\degree$. They contain a mixture of large-scale lobe-dominated sources with structures of typical sizes >100~kpc, corresponding to 30" which is resolved out on all the long baselines, together with more compact sources such as CSS \citep{2020arXiv200902750O} and a few flat-spectrum quasars (e.g. 3C345, 3C454.3). Because of the LBCS selection criteria, we expect to see a fraction of 3CRR sources in the LBCS catalogue. We find that there are 82 3CRR sources that were observed in LBCS.

\begin{figure*}
\centering
\includegraphics[scale=0.6]{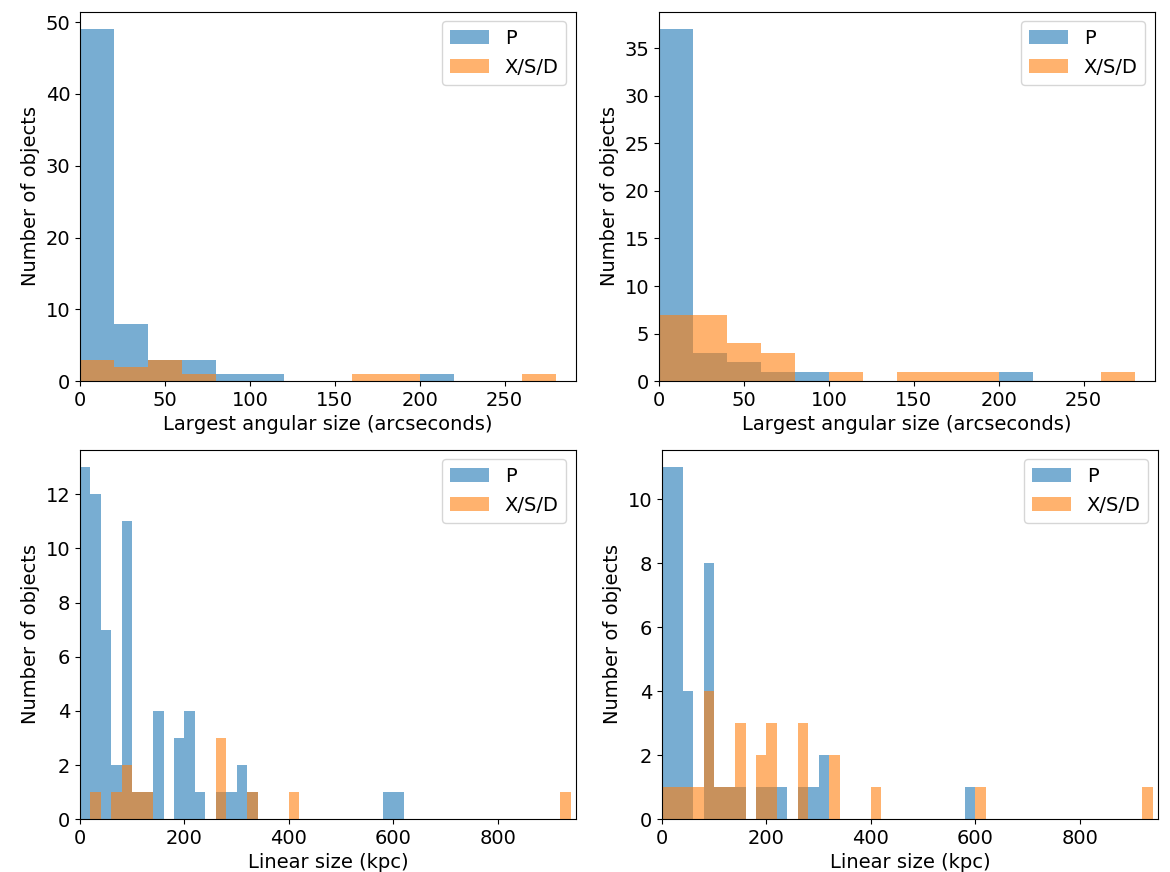}
\caption{Distribution of the largest angular sizes and linear sizes of 3CRR sources in LBCS for baselines to stations DE605/J\"ulich (left panels) and FR606/Nan\c{c}ay (right panels).}
\label{d_las}
\end{figure*}

We use previous compilations of largest angular and linear sizes in 3CRR (\url{https://astroherzberg.org/people/chris-willott/research/3crr/} in the form of an updated version of the sample presented in \citet{1983MNRAS.204..151L}). Figure~\ref{d_las} shows the relation of size to LBCS type (P, S, or X) for baselines to stations DE605/J\"ulich and FR606/Nan\c{c}ay. As expected, there is evidence that the largest sources, which are    on average thought to have less flux density in compact structure, are generally less consistently detected by LBCS, particularly on the longer baselines. 

\begin{figure}
\includegraphics[width=8cm]{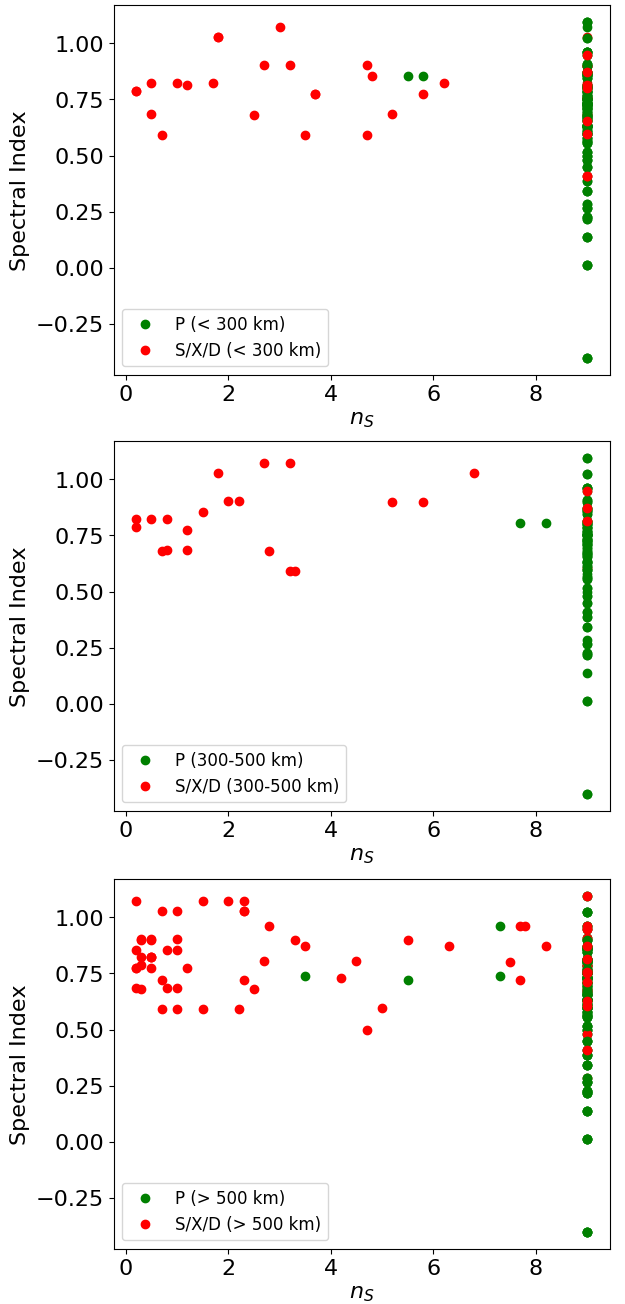}

\caption{Distribution of spectral index at (rest frame) 151 MHz vs the signal-to-noise parameter $n_S$ of LBCS sources on stations with baseline lengths to ST001 of < 300km (top), 300-500 km (centre), and > 500 km (bottom). The baseline lengths are DE605 (226 km), DE609 (227 km), DE601 (266 km), DE603 (396 km), DE604 (419 km), DE602 (581 km), SE607 (594 km), UK608 (602 km), and FR606 (700 km).} 
\label{specind}
\end{figure}

In addition to correlations with size, sources are     more likely to be detected if they have flatter low-frequency spectral indexes;  for such sources more flux density should appear at smaller spatial scales corresponding to core--jet emission rather than extended steep spectrum lobe emission. This appears to be the case (Fig.~\ref{specind}). 

\subsection{Quasars and radio galaxies}

\begin{figure}
\includegraphics[width=9.5cm]{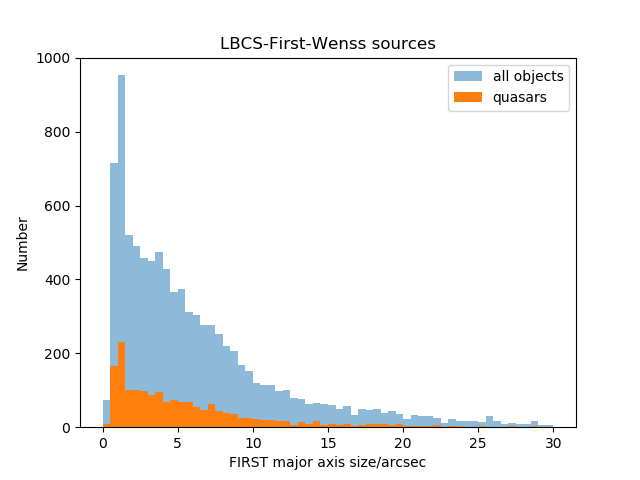}
\caption{Proportion of LBCS-WENSS-FIRST coincidences identified as quasars in {\tt milliquas} (q or Q in descriptor) compared to the statistics for all objects regardless of description.}
\label{quasars}
\end{figure}

We can also use the information about compactness provided by  LBCS  to investigate differences in the strength of flux density in compact components separately for sources identified as quasars and those identified in radio galaxies. To this end, we take the sample of LBCS sources in the overlapping WENSS/FIRST footprint and compare the catalogue with the {\tt milliquas} compilation \citep{2019arXiv191205614F}; 19\% of the sources are listed as quasars in this catalogue, the proportion of which decreases slightly with FIRST major axis size (Fig.~\ref{quasars}).

The difference between quasars and radio galaxies has been a subject of debate for 30 years, with at least some part of the answer probably due to orientation effects \citep{1987slrs.work..104S,1987ASIC..208..185P,1989ApJ...336..606B}. Radio-loud active galactic nuclei in general contain relativistically moving radio-emitting plasma in the centre, with emitted radiation having a flat spectrum, together with larger-scale jets and steep-spectrum lobe emission spread over kiloparsecs. If the jet axis is close to the line of sight to the observer, the  core--jet regions become brighter due to Doppler boosting \citep{1978PhyS...17..265B,1982MNRAS.200.1067O}, and the central regions, otherwise hidden by a dust--molecular torus, become visible. Since these central regions emit broad optical emission lines, the object is then classified as a quasar. For a given intrinsic radio power, quasars should then have brighter compact radio components than radio galaxies, as well as smaller linear size. Within the sample of quasars, flat-spectrum quasars, oriented close to the line of sight and dominated by the relativistically boosted core--jet emission, should have smaller projected linear sizes than steep-spectrum quasars, although at low frequencies we   expect the emission to be dominated by the lobes for all but a small number of sources. Earlier LOFAR studies in the Bootes field \citep{2017MNRAS.469.1883M} yielded consistency with unified schemes in that linear sizes of radio galaxies exceeded those of quasars, contrary to earlier results \citep{2014AJ....148...16S}.

\begin{figure}
\includegraphics[width=9.5cm]{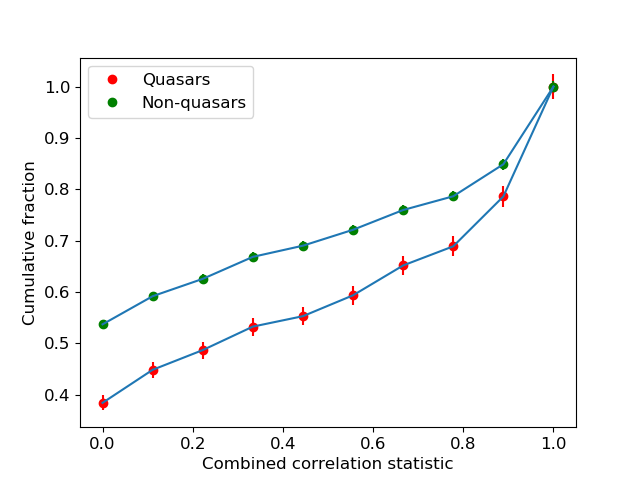}
\caption{Coherence of quasars and non-quasars, in the form of the cumulative fraction with respect to the signal-to-noise statistic of detection of correlated flux in LBCS (see text).}
\label{quascoh}
\end{figure}

For LBCS we   expect   in general that sources identified as quasars should have higher fluxes in compact structure on the 300 mas scale. This turns out to be the case. We define a combined correlation statistic as the fraction of stations for which a source is seen as P and plot this as a cumulative distribution for quasars and non-quasars (Fig.~\ref{quascoh}).

For a station close to the core (DE601/Effelsberg) we find a 55\% fraction of quasars identified as P, with significant compact radio flux density, but a 40\% fraction of non-quasars. For stations further from the core, we also obtain this difference, but a lower fraction overall (34\% P for quasars and 24\% for non-quasars). Quasars, however, are generally brighter overall because of a significant number of quasars with extrapolated 140 MHz flux density $>2$~Jy. If we exclude them the median flux for quasars and non-quasars is virtually equal, but the compact fraction is still different (50\% and 37\% respectively for quasars and non-quasars  for DE601, and 30\% and 22\% for UK608/Chilbolton), so our results here are consistent with unification.

These results are also consistent with the findings of \citet{2014AJ....147...14D} in studies of FIRST sources at milliarcsecond resolution and higher frequency (1.4~GHz). Here 30-35\% of sources identified as quasar-like in the Sloan Digital Sky Survey were found to be very compact, with a lower fraction for non-quasars. This lower fraction for non-quasars was found to increase at low flux density levels, which  are well below the $\sim$100 mJy limit of LBCS detections.

\subsection{Comparison with MWA interplanetary scintillation observations}
The Murchison Widefield Array \citep[MWA;][]{2013PASA...30....7T} is a low-frequency radio telescope located in the southern hemisphere in Western Australia at a latitude of 26.7\degr\ south.
While it shares many design characteristics with LOFAR, 128 tiles are correlated individually, providing both a wide field of view and excellent UV coverage.

The maximum baseline length for MWA Phase II is only around 6km, and so the resolution of the interferometer is only $\sim$1\arcmin.
However, \citet{2018MNRAS.473.2965M} have used it to conduct a survey of sources that show  IPS which probes similar spatial scales to the LBCS.
Most recently the Phase-II configuration \citep{2018PASA...35...33W}, which is a factor of two more sensitive for IPS observations \citep{2019PASA...36...50B}, has been used to conduct an IPS survey at 160\,MHz (Chhetri et al. in prep.).

For a detailed comparison of LBCS and the MWA IPS survey, we define a region 7hr$<RA<$11hr, +6\degree$<\delta<$+18 where both surveys have good data.
\begin{figure}
\includegraphics[width=9cm]{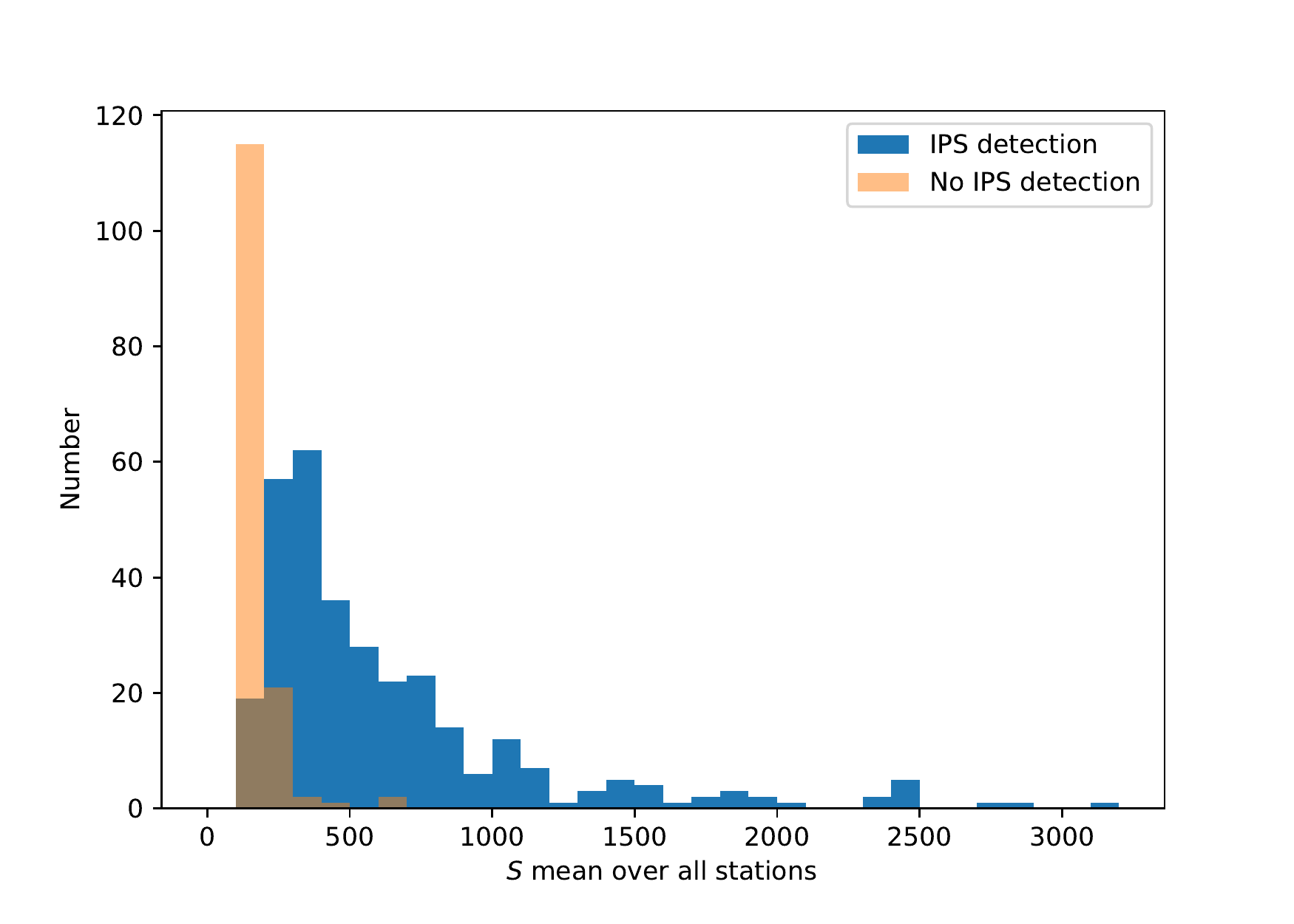}
\caption{Histogram of the mean LBCS statistic across all baselines for sources with and without IPS counterparts.}
\label{fig:ips_hist}
\end{figure}
Histograms of the mean of the $S$ statistic across all baselines are shown for all sources within this region in Fig.~\ref{fig:ips_hist}, separately for sources with and without an IPS counterpart.
IPS confirms the presence of a compact source for almost all sources with $S>$200\,mJy, the limits of which   are reliably detected by LBCS.

Interplanetary scintillation can  provide information beyond  detection versus non-detection.
\citet{2018MNRAS.474.4937C} defines the normalised scintillation index (NSI) as the ratio of the scintillating flux density to the mean flux density of the source (i.e. the source flux density at the interferometric resolution of the MWA normalised so that a point source would have a NSI of 1.
A NSI of less than one therefore indicates that a source that has some structure on scales that lie between the critical scale for IPS (approximately 0.3\arcsec) and the interferometric resolution of the MWA.
The flux density of the source on IPS scales can be calculated by multiplying the NSI by the flux density of the source measured by the MWA at the relevant frequency when used as a standard interferometer. For this we use the GLEAM survey \citep{2017MNRAS.464.1146H}.

\begin{figure}
\includegraphics[width=8cm]{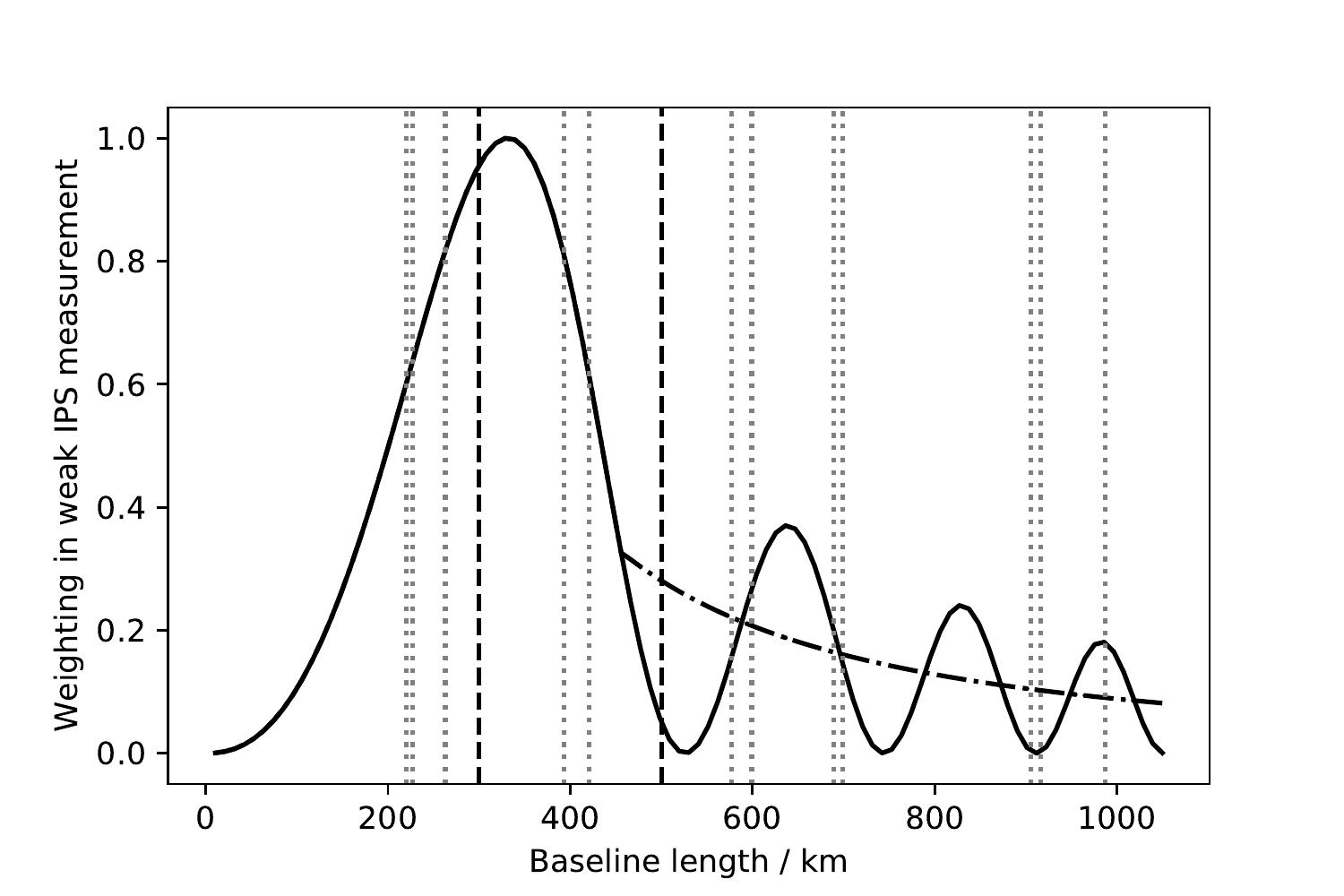}
\caption{The contribution of different spatial scales (parameterised as baseline lengths at 140~MHz) to the IPS scintillation index. Dotted vertical lines are LOFAR baselines, dashed lines delineate the 3 categories of baseline lengths defined in the text.}
\label{fig:ips_fresnel}
\end{figure}

The spatial scales that contribute to the NSI are shown in more detail in Fig.~\ref{fig:ips_fresnel}, which illustrates that they   match those probed by LBCS very closely.
Nonetheless, the measurements differ in some of the details.
The LBCS makes spot measurements of the visibility amplitude at a small number of discrete points on the UV plane (see Fig.~\ref{blength}), whereas in weak scintillation the NSI is a weighted sum of the visibility amplitude squared, where the weighting depends on the location in the visibility plane.
Following Eq. 3.2 in  \citet{1993ppl..conf..151N},  Fig.~\ref{fig:ips_fresnel} plots the weight of annuli of equal area in the UV plane. This radial symmetry is valid under the assumption of isotropy in the turbulence responsible for the scintillation in the plane of the sky; we further assume thin-screen Kolmogorov turbulence, located at a typical distance of 1\,AU.
We also plot the LOFAR baseline lengths as vertical lines, and  
to quantitatively compare the LBCS with IPS we use the following procedure.
We first split the LOFAR baselines into three categories: short ($<$300\,km), intermediate (300\,km$\le$B$<$500\,km), and long (B$\ge$500\,km).
We additionally classify Onsala as intermediate because it is the only baseline to be sufficiently foreshortened at DEC$\sim$15$\degr$ to change category.
For each LBCS source we then calculate a weight $w_i$ for the $i$th baseline such that the baselines in each category all have the same weight, and the sum of the weights of the baselines in each category have ratios 4:3:3 for short, intermediate, and long, respectively, reflecting the integral of the IPS weighting function over ranges corresponding to each of the three baseline categories.

\begin{figure}
\includegraphics[width=8cm]{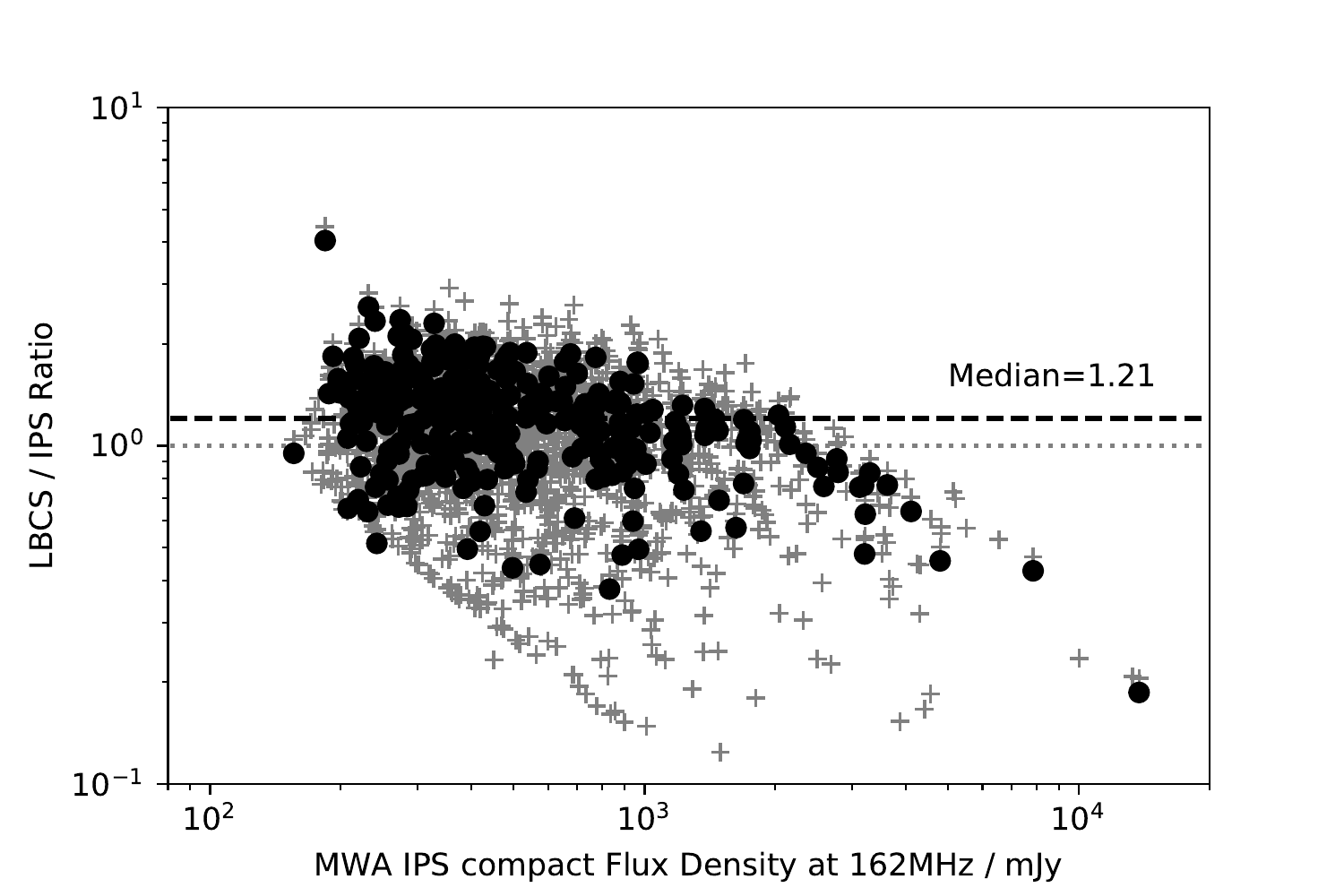}
\caption{Comparison of IPS flux densities compared with those inferred from LBCS via equation~\ref{eqn:ips_weights}. Plus signs (+)  are all LBCS sources with a counterpart; filled circles are those sources in the region 7hr$<RA<$11hr, +6\degree$<\delta<$+18 where both surveys have good data.}
\label{fig:ips_flux_ratio}
\end{figure}

The scintillation index depends on the visibility amplitude squared. Therefore, we calculate an IPS compact flux density as
\begin{equation}
        S_\textrm{IPS} = \frac{\sum_{i=0}^{i=N_{\textrm{baseline}}} w_i  S_i^2}{\sum_{i=0}^{i=N_{\textrm{baseline}}} w_i S_i}
        \label{eqn:ips_weights}
,\end{equation}
where $S_i$ is the signal strength statistic of the $i$th baseline described in Sect.~\ref{sec:catinfo}.
This can then be directly compared with the MWA NSI multiplied by the GLEAM flux density \citep{2017MNRAS.464.1146H} across the IPS observing band. 
This comparison is shown in Fig.~\ref{fig:ips_flux_ratio}, which
 shows excellent agreement for those sources in common, especially considering that LBCS is close to its southern extreme. We note that $S$ appears to saturate at approximately 2\,Jy.

\section{Conclusion}

The LOFAR Long-Baseline Calibrator Survey (LBCS) has now concluded. It covers the whole of the northern sky, and provides a calibrator network suitable for calibration of long-baseline observations with low-frequency telescopes. The calibrator density decreases at declinations below 30$^{\circ}$N, due to relative difficulty in selecting flat-spectrum candidate sources in this area of the sky and to loss of sensitivity at lower elevations. Supplementary observations may be undertaken in the future using low-frequency data at more southern declinations (e.g. \citealt{2017MNRAS.464.1146H}) and the main LoTSS survey in the north, to fill in particular gaps in the coverage. The LBCS catalogue is publicly available, and can be used to choose suitable calibrators for any observation.

We used the source statistics to investigate the nature of the LBCS calibrators.  For unresolved sources in FIRST, we find that more than 50 percent lose a significant portion of correlated flux density between scales of 2$^{\prime\prime}$ and 0\farcs 3, suggesting that they are CSS sources. Comparison with 3CRR sources shows that the largest sources are generally less detected by LBCS on the smallest scales. Those which are detected tend to have flat spectra, consistent with the dominant sources of flux density being concentrated on small scales in either core--jet or hotspot emission. Dividing all LBCS sources into quasars and non-quasars using the {\tt milliquas} database, we find that quasars are generally more compact than non-quasars, which is consistent with orientation-based unification theory. Finally, comparison with MWA allows a study of interplanetary scintillation (IPS).  There is a good agreement between the flux density on IPS scales $\gtrsim 0\farcs3$ measured by LBCS and that calculated using data from the MWA IPS survey and GLEAM. Given this good agreement, it is very likely that IPS detections can be used to select candidate calibrators with very high reliability.

In the future, development of the LOFAR-VLBI pipeline will allow for automatic generation of standardised images for calibrated LBCS sources. We hope to add these community-generated images of LBCS calibrators to the publicly available database of information. 
\begin{acknowledgements}
We thank Benito Marcote for comments on the paper. Part of this work was supported by LOFAR, the Low Frequency Array designed and constructed by ASTRON, that has facilities in several countries, that are owned by various parties (each with their own funding sources), and that are collectively operated by the International LOFAR Telescope (ILT) foundation under a joint scientific policy. This scientific work makes use of the Murchison Radio-astronomy Observatory, operated by CSIRO. We acknowledge the Wajarri Yamatji people as the traditional owners of the Observatory site. Support for the operation of the MWA is provided by the Australian Government (NCRIS), under a contract to Curtin University administered by Astronomy Australia Limited. We acknowledge the Pawsey Supercomputing Centre which is supported by the Western Australian and Australian Governments. AD acknowledges support by the BMBF Verbundforschung under the grant 052020. LKM is grateful for support from the UKRI Future Leaders Fellowship (grant MR/T042842/1). J. Mold\'on acknowledges financial support from the State Agency for Research of the Spanish MCIU through the ``Center of Excellence Severo Ochoa'' award to the Instituto de Astrof\'isica de Andaluc\'ia (SEV-2017-0709) and from the grant RTI2018-096228-B-C31 (MICIU/FEDER, EU). JPM acknowledges support from the Netherlands Organization for Scientific Research (NWO, project number 629.001.023) and the Chinese Academy of Sciences (CAS, project number 114A11KYSB20170054).

\end{acknowledgements}


\bibliographystyle{aa}
\bibliography{lobos}

\end{document}